\documentclass[aps,prd,reprint,unsortedaddress,showpacs]{revtex4-1}

\usepackage[utf8]{inputenc}
\usepackage{amssymb,amsmath}
\usepackage{ctable}
\usepackage[table]{xcolor}
\usepackage{graphicx}
\usepackage{subfigure}

\begin{document}


\title{Evidence of sub-nucleonic degrees of freedom in J/$\psi$ photoproduction in ultraperipheral collisions at the CERN Large Hadron Collider}

\author{E. Andrade-II}
\affiliation{Instituto de F\'isica da Universidade de S\~ao Paulo, SP, Brazil}
\affiliation{Texas A\&M University-Commerce, Commerce, TX, USA}
\author{I. Gonz\'alez}
\author{A. Deppman}
\affiliation{Instituto de F\'isica da Universidade de S\~ao Paulo, SP, Brasil}
\author{C. A. Bertulani}
\affiliation{Department of Physics\&Astronomy, Texas A\&M University-Commerce, Commerce, TX, USA}
\affiliation{Department of Physics\&Astronomy, Texas A\&M University, College Station, TX, USA}

\date{\today}

\begin{abstract}
We present calculations for the incoherent photoproduction of J/$\psi$ vector mesons in ultra-peripheral heavy ion collisions  
(UPC) in terms of hadronic interactions. This study was carried out using the recently developed Monte Carlo model CRISP extended to 
include UPCs at LHC energies. A careful study of re-scattering and destruction of the J/$\psi$ particles is presented for PbPb collisions at 
$\sqrt{s_{NN}} = 2.76$ TeV. We have also compared our method to AuAu collisions at $\sqrt{s_{NN}} = 200$ GeV measured at RHIC. 
\end{abstract}



\maketitle

\section{Introduction}

Photoproduction of vector mesons is important in many aspects because it provides insights into basic QCD dynamics not only in the perturbative,
but also in the non-perturbative region. The associated form factors and intermediate isobar states should test quark models. At the Large Hadron
Collider (LHC) at CERN recent experiments in pp and pPb collisions led to ions zipping past each other at relativistic energies. They are
excellent supplies of Fermi's almost real photons due to intense electromagnetic fields \cite{Fermi:1924:TSZ}, leading to numerous possibilities
for studying photonuclear and photon-photon collisions not always available with real photons
\cite{bertulani1988electromagnetic,5732455,Baltz2008}. 

Recent experiments at the LHC have reported  J/$\psi$ and $\Upsilon$  production in pp and pPb collisions in ultraperipheral collisions 
(UPC) \cite{CMS:2014ies,Abelev20131273,jeypsi}. Previous theoretical works 
have predicted the magnitude of the cross sections based on sub-nucleonic degrees of freedom 
\cite{Guzey2013,PhysRevC.84.024916,PhysRevC.85.044904,Rebyakova2012647}. One major conclusion of these efforts is that UPCs are an excellent probe of 
parton distribution functions (PDF) and the evolution of gluon distributions in nuclei \cite{PhysRevC.65.054905,Frankfurt2002220}. It is 
the goal of this work to show how a purely hadronic model could describe the incoherent photoproduction of J/$\psi$ at energies as high 
as $\sqrt{s_{NN}} = 200$ GeV and how the need for nuclear gluon dynamics at higher energies can be infered in a more reliable manner 
through the aid of an intranuclear cascade model based on hadronic consideration.

\section{Model}

Our tool for investigating  J/$\psi$ production in UPCs is the CRISP model (acronym for Collaboration Rio - Ilh\'eus - S\~ao Paulo), which 
is implemented through a cascade of hadronic collisions using Monte Carlo techniques \cite{Deppman2004,Deppman2006}. The CRISP model describes 
the nuclear reaction as a two step process, namely the intranuclear cascade and the evaporation/fission competition. For the present work the 
first one is the most important.

The intranuclear cascade encompasses all the processes from the first interaction of an incident particle with the nucleus, which is called 
primary interaction, up to the final thermalization of the nucleus \cite{Rodrigues2004}. The evaporation/fission stage describes all processes
that befall after thermalization, including all possible decay channels through strong interactions, which are successive evaporation of nucleons
or cluster of nucleons and fission \cite{Deppman2002a,Deppman2003}. Intranuclear cascade and evaporation/fission competition are also called fast
and slow processes, respectively.

A particular feature of the CRISP model is that the intranuclear cascade is described as a multicollisional process involving all nucleons in the
nuclei. This aspect allows a more realistic description of reaction mechanisms such as Pauli blocking, nuclear density fluctuations, propagation
of resonances in the nuclear medium,  final state interactions (FSI), and pre-equilibrium emissions. As a result, many different observables are
properly calculated with a small number of parameters for several nuclear masses and different collision energies.

In the evaporation/fission stage the Weisskopf mechanism for evaporation is used \cite{Weisskopf1937}, with the nuclear masses being calculated by
the Pearson nuclear mass formula \cite{Pearson2001}. The input parameters, such  as the neutron, proton and alpha particle level densities, are
calculated according to the Dostrovsky empirical formulas \cite{Dostrovsky1958}.

Both intranuclear cascade and evaporation/fission calculations with the CRISP model have been extensively investigated yielding good results for
reactions induced by photons, electrons and protons and observables such as neutron or proton multiplicity, fission and spallation cross sections,
and fragment mass distributions
\cite{Rodrigues2004,Goncalves1997,dePina1998,Deppman2002a,Deppman2002b,Deppman2004,Andrade2011,Deppman2012,Deppman2013}.

Recently the CRISP model was extended to higher energies (up to the TeV region) with the inclusion of vector meson photoproduction
\cite{Israel2014}. Some aspects of vector meson production by real photons have already been analyzed, such as subthreshold production, nuclear
transparency and FSI. In this work we apply for the first time this new tool for the study of UPC production of J/$\psi$.

The flux of virtual photons of a relativistic nucleus (projectile) is given by
\begin{equation}
 n(E_\gamma, b) = \dfrac{Z^2\alpha}{\pi^2}\dfrac{x^2}{\beta^2b^2} \left[ K_1^2(x) + \dfrac{1}{\gamma^2}K_0^2(x) \right] e^{-2\chi(b)},
 \label{eqFluxCompl}
\end{equation}
with $x = {E_\gamma b}/{\hslash\gamma \beta c}$, where $E_\gamma$ is the photon energy at the other colliding nucleus (target) frame of reference,
$b$ is the impact parameter, $Z$ is the charge of the projectile nucleus, $\alpha$ is the fine structure constant, $\beta=v/c$, $\gamma$ is the
Lorentz factor and $K_0$ and $K_1$ are the modified Bessel functions of second type. The factor $e^{-2\chi(b)}$ is the survival probability  of
both ions at impact parameter $b$, with $\chi (b)$ given by
\begin{equation}
 \chi(b) = \dfrac{\sigma_{NN}}{4\pi}\int_0^\infty dq\, q\, \tilde{\rho_t}(q) \tilde{\rho_p}(q) J_0(qb), 
 \label{chib}
\end{equation}
where $\sigma_{NN}$ is the nucleon-nucleon total cross section, $\tilde{\rho}_{t(p)}(q)$ is the Fourier transform of the nuclear density of target
(projectile) and $J_0$ is the cylindrical Bessel function of order zero. We use $\sigma_{NN} = 80$ mb for PbPb collisions and $\sigma_{NN} = 53$
mb for AuAu. We assume Fermi functions for the nuclear densities with radius $R = 6.62$ fm and diffuseness  $a = 0.546$ fm for Pb, and $R = 6.43$
fm and $a = 0.541$ fm for Au.

The cross section of a process $X$ in ultraperipheral collisions can then be calculated as \cite{bertulani1988electromagnetic}
\begin{equation}
 \sigma_X = \int \dfrac{dE_{\gamma}}{E_{\gamma}} \, N(E_{\gamma}) \, \sigma_{\gamma A \rightarrow X} (E_{\gamma}),
 \label{eqCSUPC}
\end{equation}
where $\sigma_{\gamma A \rightarrow X}$ is the cross section due to a real photon and $N(E_{\gamma})$ is the integral of $n(E_\gamma,b)$ over
impact parameters.

We represent our results  in terms of the rapidity $y$ through the relation ${d\sigma}/{dy} = E_\gamma {d\sigma}/{dE_\gamma}$, where $E_\gamma$
and the rapidity of the produced particle are related by 
\begin{equation}
 y = \ln \left[ \dfrac{W^2_{\gamma p}}{2 \gamma m_p M_P} \right] = \ln \left[ \dfrac{E_{\gamma}}{\gamma M_P} \right],
 \label{yWPhotEnRel}
\end{equation}
where $W_{\gamma p}=\sqrt{2 E_{\gamma} m_p}$ is the $\gamma p$ center-of-mass energy, $m_p$ is the proton mass, $M_P$ is the mass of the particle
of interest, and $\gamma = 2\gamma_L^2-1$ with $\gamma_L$ being the Lorentz factor of the beam in the laboratory. In terms of the rapidity of the
particle $X$ for AA collisions
\begin{equation}
 \dfrac{d\sigma_{A A \rightarrow AAX}(y)}{dy} = \dfrac{d\sigma_{\gamma A \rightarrow AX}(y)}{dy} + 
 \dfrac{d\sigma_{\gamma A \rightarrow AX}(-y)}{dy}.
 \label{diffCSyFinal}
\end{equation}

\section{Results and discussion}

\begin{figure}
 \centering
 \includegraphics[scale=0.45,keepaspectratio=true]{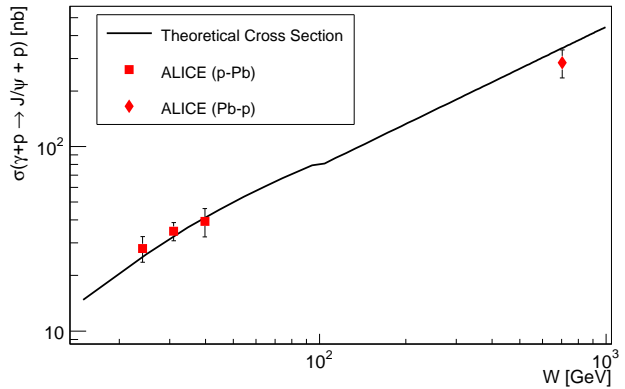}
 \caption{(Color online) J/$\psi$ photoproduction cross section off the proton. Experimental points are from the ALICE 
 Collaboration measurements in pPb UPC at $\sqrt{s_{NN}} = 5.02$ TeV \cite{ALICE2014}. }
 \label{figCSgpUltra}
\end{figure}

The CRISP model uses the universal model of soft dipole Pomeron proposed by Martynov, Predazzi and Prokudin \cite{Martynov2003,Martynov2002} to
calculate the photoproduction of meson vectors. The consistency of the model can be attested by Figure \ref{figCSgpUltra} where the cross section
for J/$\psi$ photoproduction production off the proton is compared with measurements of the ALICE collaboration for pPb ultraperipheral
collisions at $\sqrt{s_{NN}} = 5.02$ TeV \cite{ALICE2014}.

\begin{figure}
 \centering
 \includegraphics[scale=0.45,keepaspectratio=true]{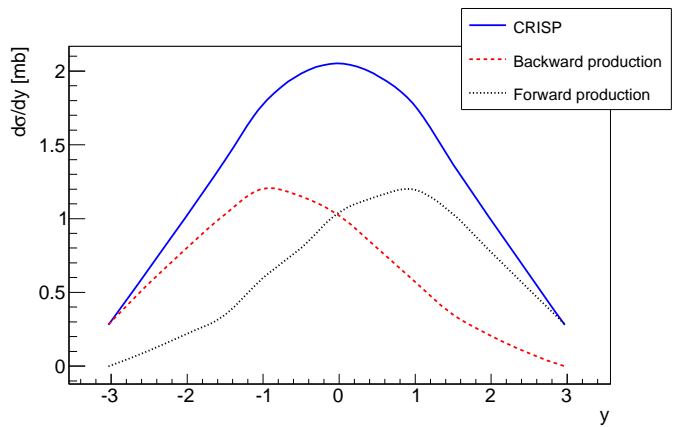}
 \caption{(Color online) Incoherent cross section for J/$\psi$ photoproduction showing contributions from both colliding 
 ions of Pb at $\sqrt{s_{NN}} = 2.76$ TeV.}
 \label{figCSJPsiExato}
\end{figure}

Experimental data on J/$\psi$ photoproduction in PbPb collisions at $\sqrt{s_{NN}} = 2.76$ TeV were published by the ALICE Collaboration
\cite{ALICE2013} where an experimental definition of coherent and incoherent production was established, according to the transverse momentum
being $p_T < 200$ MeV/c ($p_T > 200$ MeV/c) in the di-muon decay channel and $p_T < 300$ MeV/c ($p_T > 300$ MeV/c) in the di-electron decay
channel in coherent (incoherent) events. The  CRISP model yields the results displayed  in Figure \ref{figCSJPsiExato}.

We have also calculated J/$\psi$ production for several values of rapidity in the interval $-3 < y < 3$, corresponding to the range 
$219 \text{GeV} < E_{\gamma} < 89 \text{ TeV}$ for the photon energy and $20 \text{ GeV} < W_{\gamma p} < 409 \text{ GeV}$ for the $\gamma p$
center-of-mass frame. This is shown by the red (online) dashed curve in Figure \ref{figCSJPsiExato}, whereas the black (online) dotted curve is
obtained by the inversion symmetry $y\rightarrow -y$. It is readily noticed that J/$\psi$ photoproduction is dominant at lower energies because
the virtual photon flux falls rapidly with energy.

\begin{figure}
 \centering
 \includegraphics[scale=0.45,keepaspectratio=true]{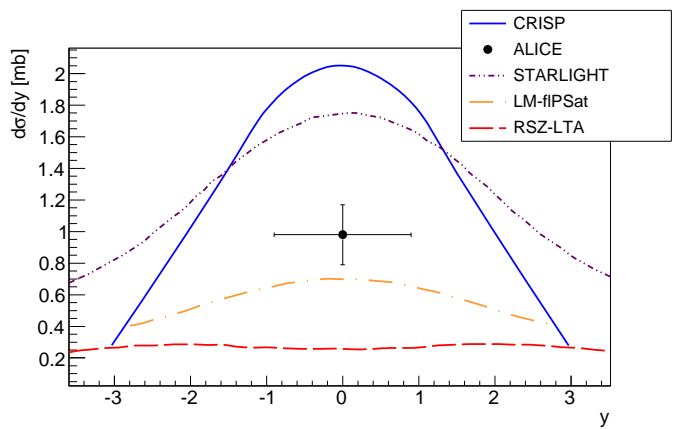}
 \caption{(Color online) Incoherent J/$\psi$ photoproduction cross section from PbPb collision at $\sqrt{s_{NN}} = 2.76$ TeV. 
 Comparison with different models and experimental data, all extracted from Reference \cite{ALICE2013}.}
 \label{figCSJPsiCompModels}
\end{figure}

Figure \ref{figCSJPsiCompModels}  compares our results with the experimental data together with results from other models, all extracted from
Reference \cite{ALICE2013}. STARLIGHT is based on a Glauber model for participating nucleons folded with the J/$\psi$-nucleon cross section and
accounting for the nuclear collision geometry \cite{Klein1999}. LM-fiPsat adopts an impact parameter saturated dipole model with an eikonalized
DGLAP-evolved gluon distribution \cite{Lappi2013}. RSZ-LTA is a partonic model in which the cross section depends on the square of the nuclear
gluon distribution.

From Figure \ref{figCSJPsiCompModels} we notice that CRISP overestimates the experimental value by nearly $100\%$. Because we use a consistent
$\gamma p$ cross section, a realistic intranuclear cascade model and proper in-medium final state interactions, such a discrepancy could be an
evidence of the limitation of a purely hadronic model, at least for this particular system. It is worthwhile mentioning that other models are not
successful either: a possible conclusion that incoherent processes in the TeV range are not very well understood.

\begin{figure}[!ht]
 \centering
 \subfigure[]{
  \includegraphics[scale=0.43,keepaspectratio=true]{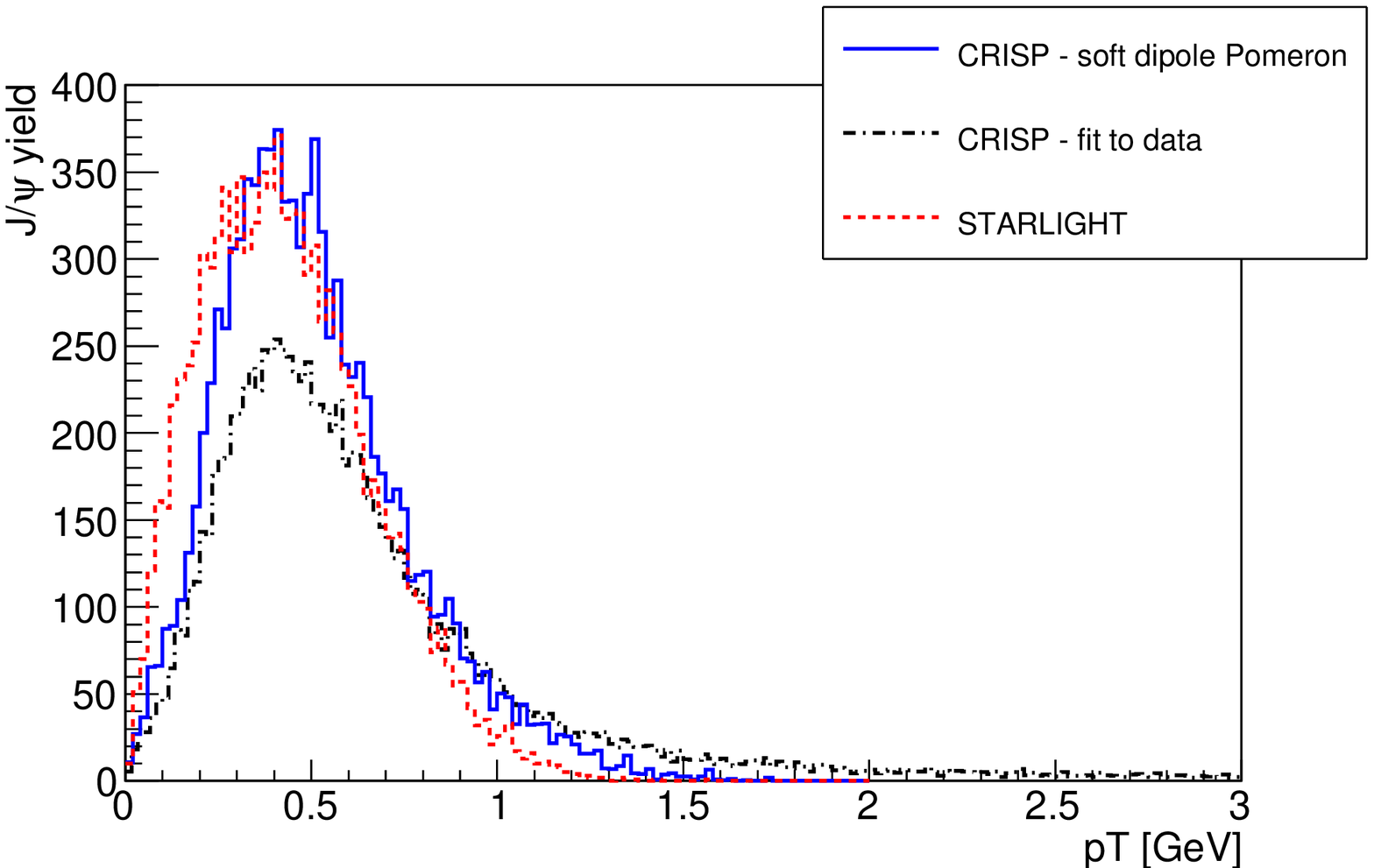}
  \label{figpTDistComp:a}
 }
 \subfigure[]{
  \includegraphics[scale=0.43,keepaspectratio=true]{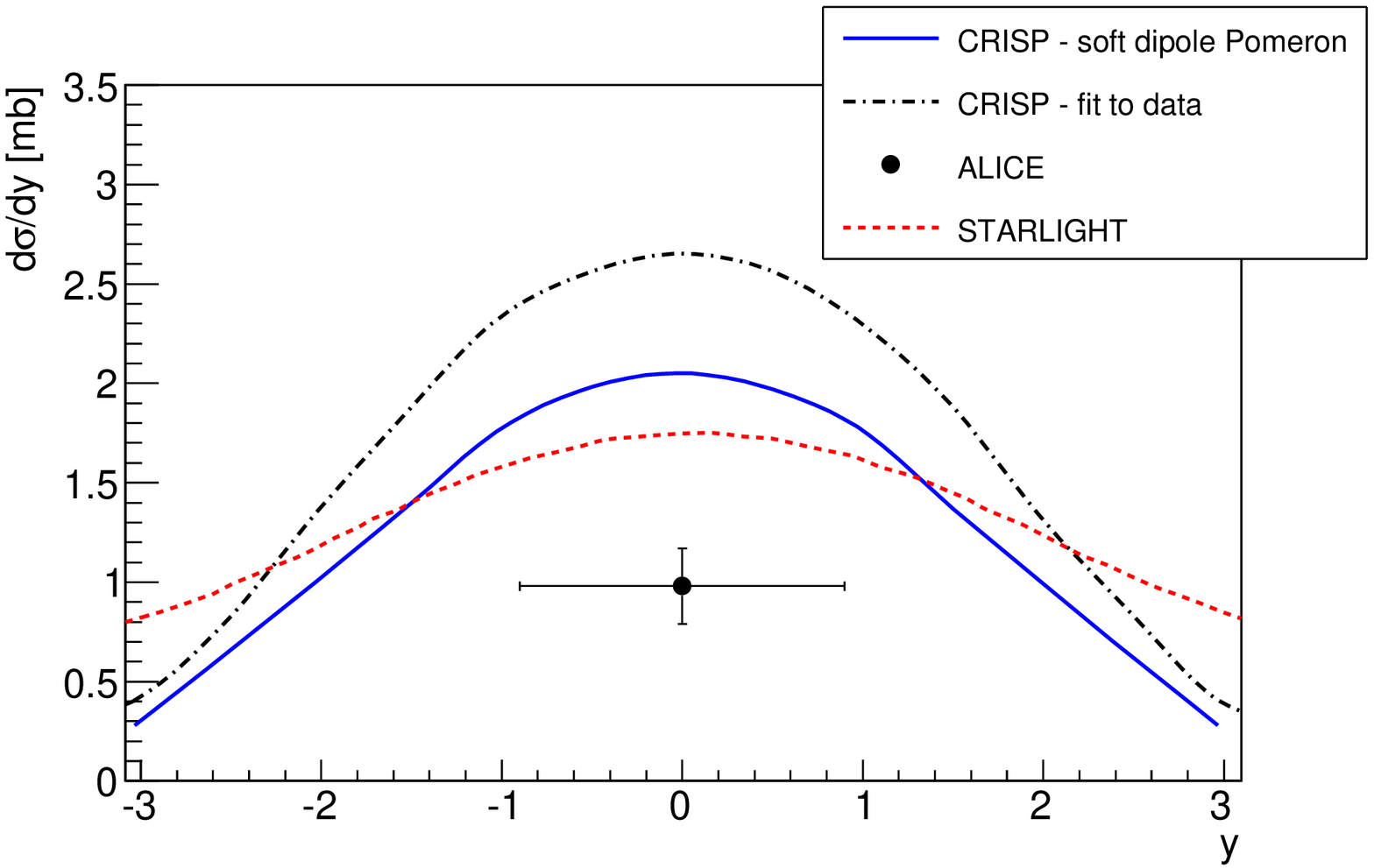}
  \label{figpTDistComp:b}
 }
 \caption{(Color online) a) Transverse momentum distributions of J/$\psi$ calculated with the CRISP model 
 compared with STARLIGHT. b) Incoherent cross section of J/$\psi$ production. PbPb collision at $\sqrt{s_{NN}} = 2.76$ TeV.}
 \label{figpTDistComp} 
\end{figure}

Transverse momentum distributions are another tool of relevance. A particular aspect of $p_T$ distributions is its sensitivity to different models
for the elastic channel of the final state interaction. Although the distributions of incoherent and coherent processes are not experimentally
accessible, a comparison between different models is nonetheless useful. Figure \ref{figpTDistComp} shows a comparison between CRISP and STARLIGHT
incoherent calculations along with the corresponding rapidity distribution. Two elastic FSI channels are provided with CRISP.

CRISP uses the Martynov, Predazzi and Prokudin soft dipole Pomeron model to evaluate the elastic FSI channel, identified as the solid line in
Figure \ref{figpTDistComp:a}. Another alternative \cite{Sibirtsev2001} is a fit to the experimental photoproduction data as
\begin{align}
 \dfrac{d\sigma}{dt}\bigg|_{t=0} = 23.15s^{0.16} + 0.034s^{0.88} + 1.49s^{0.52}, 
 \label{eqCSElastFit}
\end{align}
where $s$ is the $\gamma p$ center-of-mass energy. The first term accounts for the soft Pomeron contribution, the second one for the hard Pomeron
and the third one is the interference between the two. Eq. \eqref{eqCSElastFit} yields higher values for the transverse momenta and considerably
higher cross sections, shown in Figure \ref{figpTDistComp} by the point-dashed line. The soft dipole Pomeron model, on the other hand, is
effective in describing photoproduction off protons from threshold to several hundreds of GeV \cite{Martynov2003}. When applied to elastic FSI, it
provides a lower average transverse momemtum and the photoproduction cross section is closer to the experimental value.

Figure \ref{figpTDistComp} also shows that the $p_T$ distributions obtained with CRISP (soft dipole Pomeron) and STARLIGHT are compatible except
for two aspects. The first one is the little shift to higher momenta given by CRISP model. The second is the narrowing of the CRISP distribution
compared to STARLIGHT. This is the immediate reason why CRISP incoherent cross section is higher at $y=0$ and narrower. The differences being
considerable in terms of the physics in the models but little in terms of $p_T$ distribution. Both models agree that the mechanism called
incoherent is not sufficient to explain the data and that a different process, namely the coherent interaction, is necessary to explain the
low transverse momentum observed experimentally \cite{ALICE2013}.

\begin{figure}
 \centering
 \includegraphics[scale=0.45,keepaspectratio=true]{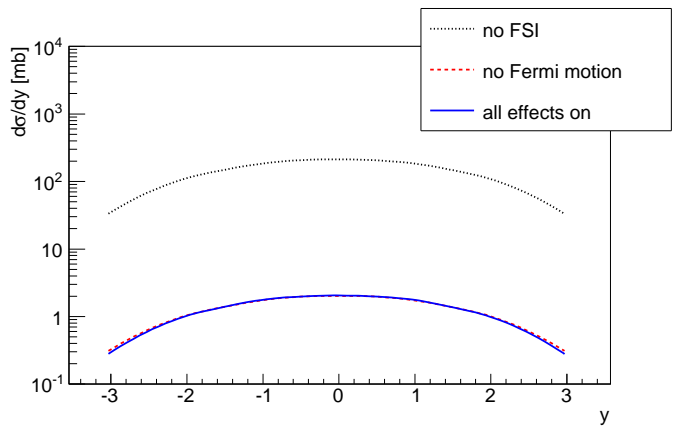}
 \caption{(Color online) Incoherent J/$\psi$ photoproduction cross section for different scenarios. PbPb collision at $\sqrt{s_{NN}} = 2.76$ TeV.}
 \label{figCSJPsiComp}
\end{figure}

\begin{figure}
 \centering
 \includegraphics[scale=0.45,keepaspectratio=true]{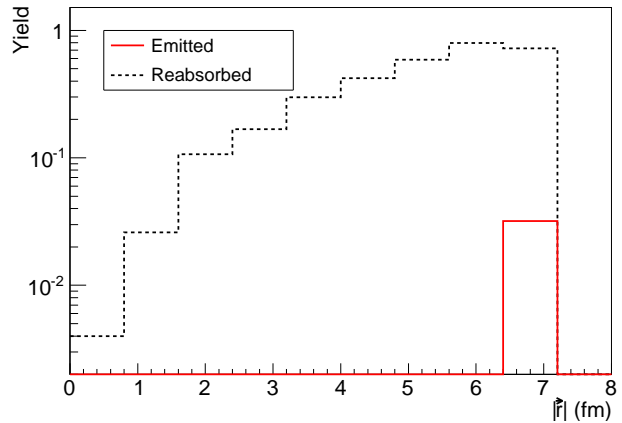}
 \caption{Distribution of the position of the created J/$\psi$ according to its emission or suppression in the nuclear matter.
 PbPb collision at $\sqrt{s_{NN}} = 2.76$ TeV.}
 \label{figPosiCriaFSION}
\end{figure}

Our model also allows to assess  nuclear medium effects  such as Fermi motion and final state interactions of J/$\psi$ in the nuclear matter. The
behavior of the cross section by switching off each of these effects can be seen in Figure \ref{figCSJPsiComp}. The cross section increases by
orders of magnitude in the absence of FSI, revealing the importance that this feature has over the results. In fact, Figure \ref{figPosiCriaFSION}
shows the position distribution of the created J/$\psi$ according to subsequent emission or suppression by the nuclear matter, evidencing that the
emitted particles are indeed those produced very close to the surface.

Two important features of the production process can be learned from Figure \ref{figPosiCriaFSION}. The large quenching of J/$\psi$ production by
FSI and the shadowing effects are noticeable due the small number of produced J/$\psi$ close to the center of the nucleus. According to a previous
work \cite{Israel2014}, the hadronization of the photon accounts for a dump in the photoproduction cross section in the internal region of the
nucleus, resulting from the shadowing effect. As a consequence the cross section is not proportional to the number of nucleons, but to $A^\alpha$,
$\alpha$ being an exponent smaller than unit. For strongly interacting particles $\alpha \approx 2/3$, meaning that the particles interact already
at the nuclear surface. However, for J/$\psi$ $\alpha \approx 0.94$, similar to values found for the photoproduction of other mesons, $\alpha
\approx 0.9$ \cite{Israel2014}. The fact that J/$\psi$ is produced also for small values of $r$ is in accordance with the shadowing effect
predictions.

The second aspect of J/$\psi$ photonuclear production is the fact that the strong final state interaction (FSI) inhibit the escape of J/$\psi$
generated in the interior of the nucleus and only those produced near the nuclear surface will escape, the others being reabsorbed by the nucleus.
In fact, as shown in Figure \ref{figCSJPsiComp}, only $\sim 1\%$ of the produced J/$\psi$ leave the nucleus. These results evidence the important
role played by FSI in the J/$\psi$ production in UPC. The effects of FSI can be investigated in LHC energies using UPCs if collision between
nucleus of different sizes as CC and UU (besides pA collisions) can be performed.

With respect to Fermi motion, we observe that its effects are relatively small compared to the full calculation, of the order of the uncertainties
in the Monte Carlo method.  

\begin{figure}[!ht]
 \centering
 \includegraphics[scale=0.45,keepaspectratio=true]{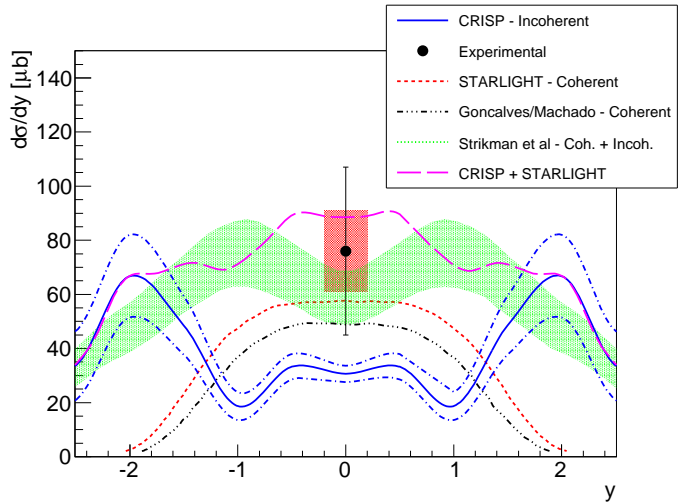}
 \caption{Total photoproduction cross section of J/$\psi$ in AuAu collision at $\sqrt{s_{NN}} = 200$ GeV. Comparison with STARLIGHT, Strikman
 \textit{et al} and Gon\c calves-Machado models, all extracted from \cite{PHENIX2009} as well as the experimental data. The lines 
 (\textendash \textperiodcentered) delimit the statistical uncertainties in the calculations.}
 \label{figCSJPsiExatoAu_Models}
\end{figure}

J/$\psi$ photoproduction in AuAu collisions at $\sqrt{s_{NN}} = 200$ GeV was also calculated in the $-2 < y < 2$ rapidity interval corresponding
to the range $45 \text{ GeV} < E_{\gamma} < 2.44 \text{ TeV}$ for the photon energy and $9 \text{ GeV} < W_{\gamma p} < 68 \text{ GeV}$. The
calculation is shown in Figure \ref{figCSJPsiExatoAu_Models} compared with experimental data from the PHENIX Collaboration \cite{PHENIX2009}. The
data corresponds to the total cross section measured without separation between coherent and incoherent events due to statistical limitations. 

In this case the PHENIX collaboration estimated a dominant coherent contribution, the reason why comparisons with coherent calculations were
provided in Reference \cite{PHENIX2009}. STARLIGHT calculations and the Gon\c calves-Machado model were extracted from Ref. \cite{PHENIX2009}. The
calculations provided by Strikman \textit{et al} for the total cross section is also extracted from Ref. \cite{PHENIX2009}.

According to estimates by the PHENIX Collaboration, the incoherent contribution is approximately $40\%$, or $\sim 30\, \mu b$, in accordance with
CRISP model calculations. Figure \ref{figCSJPsiExatoAu_Models} also shows the summation of CRISP and STARLIGHT results. A fair agreement is found
between the summation and both Strikman model and the experimental data. The lack of more experimental data certainly reduces the extent of the
analysis.

The slight disagreement between CRISP  calculation and experimental results for PbPb at $\sqrt{s_{NN}} = 2.76$ TeV cannot be attributed to FSI.
We have verified that the effects of FSI saturate since the J/$\psi$ effectivelly produced in UPC are those generated exactly at the nuclear 
surface, so small modifications on J/$\psi$ FSI will not alter our results. We have also tested possible effects of a smooth surface by modifying 
the parameters of the survival probability inside a reasonable range. The decrease of the cross section for $y=0$ is smaller than $6$\% and 
thus our conclusions are not modified.

The disagreement between calculation and experiment therefore can be attributed only to the primary interaction between the virtual photon and
the nucleon. Since the model used is a nucleonic one, this can be an indication of the necessity of sub-nucleonic degrees of freedom in the 
description of J/$\psi$ photoproduction. In this case the better agreement obtained with AuAu collision could be indicative of the fact that
at lower energies the nucleon-based model is still satisfactory. In fact, considering again $y=0$, the photon energy at target reference frame
for Pb and Au are respectively $E_{\gamma} = 4.4$ TeV, $0.3$ TeV. The threshold for sub-nucleonic degrees of freedom would be inside this interval.

The present analysis is limited by the lack of experimental data. It would be interesting to have the knowledge of J/$\psi$ production in UPC 
for both Pb and Au at different energies in the interval $\sqrt{s_{NN}} = 200$ GeV to $2.76$ TeV.

\section{Conclusions}

In summary, the J/$\psi$ photoproduction  for PbPb collisions at $\sqrt{s_{NN}} = 2.76$ TeV and for AuAu collisions at $\sqrt{s_{NN}} = 200$ GeV
was studied with the CRISP model and compared with the existing experimental data and models. The CRISP hadronic model describes reasonably the
photoproduction of J/$\psi$ in UPCs at lower energies ($\leq 200$ GeV) but with limitations at higher energies. An advantage and partial success
of the model is the use of the correct photoabsorption cross sections for the different channels of sequential hadronic collisions  with the final
state interactions of the J/$\psi$  proven to be of great relevance. Our findings led to the reliable conclusion that the inclusion of 
sub-nucleonic degrees of freedom is needed to describe J/$\Psi$ photproduction in UPCs.

\begin{acknowledgments}
 We acknowledge the support from the Brazilian agency FAPESP under grant 2012/13337-0, the U.S. NSF Grant No. 1415656, and U.S. DOE grant No. 
 DE-FG02-08ER41533. Also, we thank Dr. J. D. Tapia Takaki for his valuable considerations on this study, greatly improving our analysis.  
\end{acknowledgments}


\end{document}